\documentclass[letterpaper,10pt,conference]{ieeeconf}
\pdfoutput=1

\usepackage{color}
\usepackage{times}
\usepackage[dvipsnames]{xcolor}
\usepackage{amsmath, mathtools}
\usepackage{amssymb}
\usepackage{graphicx}
\usepackage{epsfig}
\usepackage{booktabs}
\usepackage{bm}
\usepackage{url}
\usepackage[font=footnotesize, skip=4pt]{caption}
\usepackage{subfig}
\usepackage{amsthm}

\newtheorem{definition}{Definition}

\newtheorem{theorem}{Theorem}
\newtheorem{remark}{Remark}

\newtheorem{assumption}{Assumption}

\IEEEoverridecommandlockouts
\usepackage{tikz}
\usetikzlibrary{shapes,arrows}
\usetikzlibrary{bayesnet}
\usetikzlibrary{positioning}
\tikzstyle{status} = [rectangle, draw=black, text centered, anchor=north, text=black, minimum width=9em, minimum height=3em, node distance=6ex and 7em]
\tikzstyle{line} = [draw,thick,-latex]
\tikzstyle{transition} = [font=\small]

\makeatletter
\newcommand{\revisionhistory}[1]{%
\@ifundefined{showrevisionhistory}{\relax}{%
{#1}%
}%
}
\makeatother

\begin{document}


\title{\LARGE \bf Eliciting Private User Information for Residential Demand Response}
\author{Datong P. Zhou$^{\ast\dagger}$, Maximilian Balandat$^\star$, Munther A. Dahleh$^\dagger$, and Claire J. Tomlin$^\star$
\thanks{$^\ast$Department of Mechanical Engineering, University of California, Berkeley, USA. {\tt\footnotesize datong.zhou@berkeley.edu}}
\thanks{$^\dagger$Laboratory for Information and Decision Systems, MIT, Cambridge, USA.
{\tt\footnotesize [datong, dahleh]@mit.edu}}
\thanks{$^\star$Department of Electrical Engineering and Computer Sciences, University of California, Berkeley, USA.
{\tt\footnotesize [balandat, tomlin]@eecs.berkeley.edu}}%
\thanks{This work has been supported in part by the National Science Foundation under CPS:FORCES (CNS-1239166) and CEC Grant 15-083.}
}

\maketitle
\thispagestyle{empty}
\pagestyle{empty}

\begin{abstract}
Residential Demand Response has emerged as a viable tool to alleviate supply and demand imbalances of electricity during times when the electric grid is strained. Demand Response providers bid reduction capacity into the wholesale electricity market by asking their customers under contract to temporarily reduce their consumption in exchange for a monetary incentive. This paper models consumer behavior in response to such incentives by formulating Demand Response in a Mechanism Design framework. In this auction setting, the Demand Response Provider collects the price elasticities of demand as bids from its rational, profit-maximizing customers, which allows targeting only the users most susceptible to incentives such that an aggregate reduction target is reached in expectation. We measure reductions by comparing the materialized consumption to the projected consumption, which we model as the ``10-in-10''-baseline, the regulatory standard set by the California Independent System Operator. Due to the suboptimal performance of this baseline, we show, using consumption data of residential customers in California, that Demand Response Providers receive payments for ``virtual reductions'', which exist due to the inaccuracies of the baseline rather than actual reductions. Improving the accuracy of the baseline diminishes the contribution of these virtual reductions.
\end{abstract}


%
%


\section{Introduction}
\label{sec:Introduction}
With the restructuring of the traditional, vertically integrated energy market towards a competitive market, Demand-Side Management (DSM) has become a viable tool for alleviating supply and demand imbalances of electricity. Facilitated by advancements in information and communications technology, smart metering infrastructure allows end-users of electricity to ``participate'' in the electric market as virtual power plants through properly designed incentive mechanisms. DSM is motivated by the inelasticity of energy supply, which causes small variations in demand to result in a price boom or bust, respectively. These price fluctuations are aggravated by the inherent volatility of renewable generation resources, their increasing levels of penetration, and the prohibitively high capital cost of energy storage. Since a load-serving entity (LSE) is required to procure electricity at fluctuating prices to cover the electricity demand of its residential households under contract instantaneously and at quasi-fixed tariffs, price risks are almost entirely borne by the LSE. Incentivizing users to temporarily reduce their consumption (and charging a fee if users do not reduce) during periods of high prices therefore partially passes such price risks on to customers.

While the area of DSM has attracted a vast array of research across different domains (see \cite{Palensky:2011aa} for a summary), we in this paper focus on the area of Demand Response (DR), where end-users of electricity are incentivized to reduce their demand temporarily during designated hours, precisely when there is a shortage of electricity supply. Users receive a reward for each unit of reduction, but incur a penalty for increasing their consumption. Demand Response providers (DRPs) bundle these reductions and can offer these reductions as a bid directly into the competitive wholesale electricity market, or enter bilateral contracts with load-serving utilities. While DR is traditionally carried out on a commercial level, residential customers are targeted for load reduction programs, as well. For instance, in California, the Public Utilities Commission (CPUC) launched a ``Demand Response Auction Mechanism'' (DRAM) in July 2015 \cite{State-of-California:aa} to allow DRPs to offer reduction capacity from residential customers directly into the day-ahead electricity market, where they are subject to regular market clearing prices and shortfall penalties. Utilities are required to purchase a fixed minimum monthly amount of this reduction capacity.

To make an informed capacity bid into the market, the DRP must take various factors into account, such as the expected Locational Marginal Price (LMP) which determines its market clearing price, the elasticity of users' demand given an incentive, and the number of users under contract. If the DRP bids too much capacity, the aggregate reduction among its user base will likely fail to reach the capacity volume, thereby incurring a shortfall penalty; similarly, a suboptimal revenue arises from too small a bid. The DRP can improve its bidding strategy by learning users' behavior in response to incentives. However, users' preferences are typically private information and hence unknown to the utility. The challenge thus becomes to elicit this private information. We cast this problem as a mechanism design problem, where the DRP as the auctioneer solicits bids from each of its residential customers through an \textit{incentive compatible} and \textit{individually rational} mechanism. The motivation behind this approach is to increase allocative efficiency, that is, the utility would like to solicit reductions only from the highest reducers, who are most willing to reduce their consumption in exchange for the lowest possible reward. In this paper, we design such a mechanism that fulfills these criteria and benchmark its performance against the omniscient case, where user characteristics are common knowledge.

A crucial question that arises from this setting is how to measure the reduction of any individual user during a DR event, given that only the consumption outcome under a treatment can be observed, but not its counterfactual (the consumption had there been no DR event). This is the fundamental problem of causal inference \cite{Holland:1986aa}. To estimate the reduction during any particular DR event, it is thus essential to estimate the counterfactual, which we refer to in this context as ``baseline''. Estimating this baseline in the absence of a Randomized Controlled Trial is a modern area of research at the intersection of economics and machine learning. Examples for such baseline estimates can be found in \cite{Athey:2016aa, Abadie:2012aa, Zhou:2016aa, Zhou:2016ab}. In this paper, we employ the ``10-in-10'' baseline employed by the California Independent System Operator (CAISO) \cite{CAISO}, which estimates the counterfactual for a particular DR event as the mean consumption of the 10 most recent days during the same hour as the DR event. Using this baseline, the measured reduction for any selected user can be formulated as the sum of a virtual reduction, which reflects the estimation error in the baseline prediction, and the actual reduction due to price elasticity of user demand. We observe that the DR provider can achieve a virtual reduction from those users for which the baseline is high. That is, the DR provider receives payments for virtual, non-existent reductions which are indirectly paid for by utilities. However, we show that a more accurate baseline diminishes the impact of such virtual reductions.

\subsection*{Related Work}
Modeling consumer behavior in response to monetary incentives in DR and their heterogeneity is a growing area of research. In \cite{Kwac:2013aa}, the authors formulate the problem of targeting the ``right'' customers for DR as a stochastic knapsack problem in order to achieve a target reduction with high probability. However, users' responses are modeled as a linear model without private user information.

Other works have incorporated a contractual formulation between consumers and suppliers in DR settings. For example, \cite{Li:2015aa} designs a DR market where suppliers bid supply curves in the presence of a supply shortage to the load-serving entity and analyzes the ensuing market equilibria. In \cite{Balandat:2014aa}, the authors formulate a contract between an aggregator of buildings, individual buildings, and the wholesale electricity market to exploit flexibility of commercial buildings' HVAC consumption. In a similar fashion, \cite{Han:2010aa} formulates a contract design problem between an aggregator and individual electric vehicle owners to maximize its revenue by providing power capacity to the grid operator.

To quantify the impact of DR signals on the reduction of consumption, \cite{Zhou:2016aa, Zhou:2016ab} estimate individual treatment effects in response to hourly DR events by comparing the estimated counterfactual consumption to the actual, observed consumption. \cite{Li:2016aa} formulates an optimal treatment assignment strategy to precisely measure the treatment effect of DR.

The application of Mechanism Design on DR is covered in \cite{Samadi:2012aa}, where the authors maximize the social welfare of consumers and the energy provider by designing a consumption controller with a Vickrey-Clarke-Groves auction. In \cite{Ma:2016aa, Li:2016ab}, the authors incorporate uncertainty into consumers' reduction behavior and introduce the notion of reliability for achieving a designated amount of aggregate reduction. 

\subsection*{Contributions}
Unlike previous works, which modeled reductions as multiples of unit reductions, we account for the \textit{Fundamental Problem of Causal Inference} \cite{Holland:1986aa} into the mechanism design formulation between DRP and users, which is our main contribution. Specifically, we estimate reductions using the CAISO ``10-in-10'' baseline as the counterfactual estimate. As a consequence of uncertain baseline predictions, \textit{virtual reductions} arise. Using observational data from residential customers in California, we quantify the extent to which these virtual reductions counteract DR, and how these reductions diminish as baseline estimates become more precise.

\subsection*{Notation}
Let $[\hspace{0.05cm}\cdot\hspace{0.05cm}]_+ = \max(0, \cdot)$. Vectors are printed in boldface. Let $\mathbf{a}_{-i}$ denote the vector of all components in $\mathbf{a}$ excluding $i$. $\mathbf{1}_{(\cdot)}$ denotes the indicator function.

\subsection*{Outline}
The remainder of this paper is organized as follows: Section \ref{sec:Preliminaries} characterizes DR market participants and their interactions, based on which Section \ref{sec:Mechanisms} presents a mechanism for the DR provider to elicit private user information and to achieve an aggregate reduction among its users under contract. Section \ref{sec:baseline_gaming} elucidates the difference between virtual and actual reductions as an artifact of an uncertain baseline estimate. The mechanism is simulated on residential smart meter data in California in Section \ref{sec:Simulations}, where we experimentally show how more accurate baselines reduce the amount of virtual reductions. Section \ref{sec:Conclusion} concludes. All proofs are relegated to the Appendix.

%
%


\section{Market Participants and Interactions}
\label{sec:Preliminaries}
\subsection{Residential Demand Response}
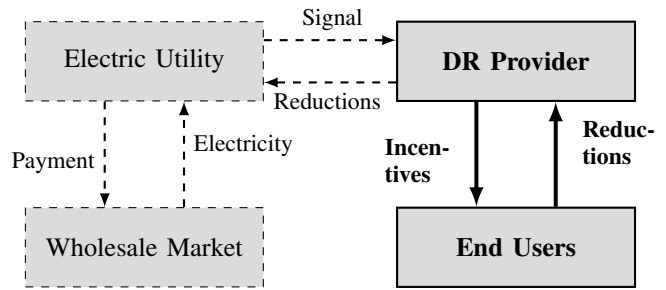
\begin{figure}[h]
\centering
\begin{tikzpicture}
\node [status, dashed, align=left, fill={rgb:black,0.5;white,3}] (Ut) {Electric Utility};
\node [status, dashed, below=4em of Ut, fill={rgb:black,0.5;white,3}] (WSM) {Wholesale Market};
\node [status, thick, right=5em of WSM, fill={rgb:black,0.5;white,3}] (EU) {\textbf{End Users}};
\node [status, thick, right=5em of Ut, fill={rgb:black,0.5;white,3}] (DRP) {\textbf{DR Provider}};
\path [line, dashed] ([xshift=-1.5em]Ut.south) -- ++(0,-0.67) -- ++(0,0) -- ([xshift=-1.5em]WSM.north) node [transition,pos=0.2,left] {Payment};
\path [line, dashed] ([xshift=1.5em]WSM.north) -- ++(0,0.67) -- ++(0,0) -- ([xshift=1.5em]Ut.south) node [transition,pos=0.2,right] {Electricity};

\path [line, line width=1.5] ([xshift=-1.5em]DRP.south) -- ++(0,-0.67) -- ++(0,0) -- ([xshift=-1.5em]EU.north) node [transition,pos=0.2,left] {\begin{tabular}{l}\textbf{Incen-}\\
\textbf{tives}\end{tabular}};
\path [line, line width=1.5] ([xshift=1.5em]EU.north) -- ++(0,0.67) -- ++(0,0) -- ([xshift=1.5em]DRP.south) node [transition,pos=0.2,right] {\begin{tabular}{l}
	\textbf{Reduc-}\\
    \textbf{tions}
\end{tabular}};

\path [line, dashed] ([yshift=0.8em]Ut.east) -- ++(0.7,0) -- ++(0,0) -- ([yshift=0.8em]DRP.west) node [transition,pos=0.2,above] {Signal};
\path [line, dashed] ([yshift=-0.8em]DRP.west) -- ++(-0.7,0) -- ++(0,0) -- ([yshift=-0.8em]Ut.east) node [transition,pos=0.2,below] {Reductions};

\end{tikzpicture}
\vspace{0.04cm}
\caption{Energy Market Participants for DR and their Interactions}
\label{fig:interaction}
\end{figure}

Figure \ref{fig:interaction} describes the interaction between the DRP, end-users, the electric utility, and the wholesale electricity market.

The DRAM requires electric utilities to acquire demand flexibility from DRPs, which they submit as part of their supply curves as a bid into the real-time wholesale electricity market. If these bids are cleared, the utility sends the DRP a signal to ask for a specified aggregate reduction among its users. The DRP elicits reductions by incentivizing a subset of its customers $\mathcal{T} \subseteq \mathcal{I}$ with user-specific per-unit rewards $\lbrace r_i\in\mathbb{R}_+~|~i\in\mathcal{T} \rbrace$, where $\mathcal{I}=\lbrace 1, \ldots, n\rbrace$ denotes the set of users. In exchange for the monetary incentive, users reduce consumption by $\lbrace \delta_i\in\mathbb{R}~|~i\in\mathcal{T}\rbrace$. A per-unit penalty $q\in\mathbb{R}_+$, which is assumed to be identical for all users and common knowledge, is enforced for an increase in consumption beyond the baseline. Users in the non-targeted group $\mathcal{I}\setminus \mathcal{T}$ are excluded from the incentive program. In this paper, we focus on the interaction between the DRP and the end-users from the perspective of the DRP. To maximize its profit, the goal of the DRP is to achieve an a-priori defined aggregate reduction with minimal payments to its users.

\subsection{Residential Customers}
Each rational, profit-maximizing user $i\in\mathcal{I}$ is endowed with the ``10-in-10'' baseline $\hat{x}_i\in\mathbb{R}_+$ employed by the California Independent System Operator (CAISO) \cite{CAISO}, which is an estimate of her counterfactual consumption \cite{PJM} for a particular hour. For notational ease, we drop time indices, but we emphasize the need to re-calculate $\hat{x}_i$ for any individual hour. The baseline for a particular hour on a weekday is calculated as the mean of the hourly consumptions on the 10 most recent business days during the hour of interest. For weekend days and holidays, the mean of the 4 most recent observations is calculated. User $i$'s measured load reduction $\delta_i$, provided she is given incentive $r_i$ to reduce during a particular hour, is simply the difference between the baseline $\hat{x}_i$ and the actual, materialized consumption $x_i$:
\begin{align}\label{eq:reduction_user}
\delta_i = \begin{cases}
0~ &, ~\text{if}\quad i \not\in \mathcal{T} \\
\hat{x}_i - x_i~ &, ~\text{if}\quad i\in\mathcal{T}
\end{cases}
\end{align}
Due to the widespread existence of advanced metering infrastructure, the baseline $\hat{x}_i$ is assumed to be common knowledge among the DRP and user $i$. The utility of user $i$ is defined as follows:
\begin{align}\label{eq:utility_under_dr_mechanism}
u_i = \begin{cases}
0 & ,~\text{if}\quad i \not\in \mathcal{T} \\
r_i\cdot[\hat{x}_i - x_i]_+ - q\cdot[x_i - \hat{x}_i]_+& ,~\text{if}\quad i\in\mathcal{T}
\end{cases}
\end{align}
which equals the payment from the DRP to user $i$. That is, if the user is under a DR contract with the DRP, she is rewarded with $r_i\in\mathbb{R}_+$ for each unit of reduction, and charged $q$ for each unit of consumption above the baseline $\hat{x}_i$.

We model users' consumption in response to $r_i$, denoted with $x_i(r_i)$, with a semi-logarithmic demand curve, an assumption frequently made in economics:
\begin{align}
x_i(r_i) &= \bar{x}_i\cdot \exp(-\alpha_i r_i)\nonumber\\
\log x_i(r_i) &= \log\bar{x}_i - \alpha_i r_i\quad\quad~\forall~i\in\mathcal{I}\label{eq:semilog_demand_curve}
\end{align}
In \eqref{eq:semilog_demand_curve}, $\bar{x}_i\in\mathbb{R}_+$ and $\alpha_i\in\mathbb{R}_+$ are random variables signifying the base demand (the intercept or the consumption with $r_i=0$) and the slope of the demand curve in log-linear coordinates, respectively. This semi-logarithmic demand curve captures the fact that the amount of reduction is marginally decreasing in the reward $r_i$ and saturates. User $i$'s type $\boldsymbol{\theta}_i$ is information correlated with $(\bar{x}_i, \alpha_i)$ (not necessarily $(\bar{x}_i, \alpha_i)$ itself) and user $i$'s private information.

\subsection{Demand Response Provider}
The DRP aims to maximize its profit $\Pi$ in expectation:
\begin{equation}
\begin{aligned}\label{eq:DRP_profit}
\Pi =&~ \bar{r}\cdot\min(\Delta, M) - \bar{q}\cdot\left[ M-\Delta \right]_+ \\
&- \sum_{i\in\mathcal{I}}\delta_i\left( r_i \cdot\mathbf{1}_{\delta_i<0} - q_i \cdot\mathbf{1}_{\delta_i\geq 0}\right).
\end{aligned}
\end{equation}
$\Pi$ is random in $\delta_1, \ldots, \delta_n$. $\Delta = \sum_{i\in\mathcal{I}}\delta_i$ is the total sum of reductions and $M\in\mathbb{R}_+$ the target capacity the DRP has to provide to the utility. $\bar{r}$ and $\bar{q}\in\mathbb{R}_+$ denote the per-unit reward and shortfall penalty the DRP is subject to in the wholesale electricity market. Note that $\bar{q}\neq q$ and $\bar{r}\neq r_i$. The first term of \eqref{eq:DRP_profit} represents the profit the DRP earns for materialized reductions, the second term captures the shortfall penalty for unfulfilled reductions, and the last term is the sum of payments disbursed to individual customers.
\begin{assumption}\label{as:DRP_risk_neutral}
The DRP is risk-neutral and profit-maximizing.
\end{assumption}
\begin{assumption}\label{as:shortfall_penalty_wholesale}
The per-unit penalty $\bar{q}$ in the wholesale electricity market and the per-unit reward $\bar{r}$ are greater than the maximum per-unit reward disbursed to any customer, i.e. $\min(\bar{q}, \bar{r}) > \max_{1\leq i\leq n}(r_i)$.
\end{assumption}
With Assumptions \ref{as:DRP_risk_neutral} and \ref{as:shortfall_penalty_wholesale}, \eqref{eq:DRP_profit} can be rewritten as follows:
\begin{equation}\label{eq:DRP_optimization_problem}
\begin{aligned}
& \underset{r_1, \ldots, r_n}{\text{minimize}}
& & \mathbb{E}_{\delta_1, \ldots, \delta_n}\left[\sum_{}\mathop{}_{\mkern-5mu i\in\mathcal{I}}\delta_i \left( r_i \mathbf{1}_{\delta_i<0} - q_i \mathbf{1}_{\delta_i\geq 0}\right)\right] \\
& \text{subject to}
& & \mathbb{E}_{\delta_1, \ldots, \delta_n}\left[\sum_{}\mathop{}_{\mkern-5mu i\in\mathcal{I}}\delta_i\right] \geq M.
\end{aligned}
\end{equation}

That is, the DRP aims to find an optimal vector of per-unit rewards $\mathbf{r}^\ast$ that minimizes the expected total amount of payments disbursed to the users while satisfying the constraint that the expected sum of reductions exceeds $M$.

%
%


\section{Demand Response Mechanism}
\label{sec:Mechanisms}
To find an approximation to the solution of \eqref{eq:DRP_optimization_problem}, the utility needs to elicit user $i$'s private type $\boldsymbol{\theta}_i$ with an incentive compatible (IC) and individually rational (IR) mechanism. IR guarantees that participation in the mechanism, provided users act rationally, results in an expected payoff that is at least as large as in the case of non-participation (outside option), which is zero in our case \eqref{eq:utility_under_dr_mechanism}. IC is required to ensure that users report their types truthfully to the DRP.

\subsection{Mechanism Design Basics}
We first introduce basic notation relevant to our problem. Let $\boldsymbol{\theta}$ denote the collection of types $(\boldsymbol{\theta}_1, \ldots, \boldsymbol{\theta}_n)$, where each $\boldsymbol{\theta}_i\in\boldsymbol{\Theta}_i~\forall~i\in\mathcal{I}$ is drawn from its type space $\boldsymbol{\Theta}_i$. It is assumed that $\boldsymbol{\theta}$ is drawn from a commonly known joint distribution $F$ defined on the product space $\boldsymbol{\Theta} = \times_{i=1}^n \boldsymbol{\Theta}_i$. Each agent is assumed to seek expected utility maximization of her utility function $u_i(\mathbf{y}, \boldsymbol{\theta}_i) : \mathcal{Y} \times \boldsymbol{\Theta}_i\mapsto \mathbb{R}$, where $\mathbf{y}=(\mathbf{d}, \mathbf{r})\in \mathcal{Y}=\lbrace 0, 1\rbrace^n \times \mathbb{R}_+^n$ is the collective choice consisting of the vector of allocation decisions $\mathbf{d}$ and the vector of rewards $\mathbf{r}$. The social choice function $f(\boldsymbol{\theta}):\boldsymbol{\Theta} \mapsto \mathcal{Y}$ maps a particular collection of types $\boldsymbol{\theta}$ to $\mathbf{y}$.

Let $\mathcal{S}_i, \ldots, \mathcal{S}_n$ denote the strategy spaces of users $i\in\mathcal{I}$. A realized strategy vector $\mathbf{s}\in \times_{i=1}^n \mathcal{S}_i$ defines an outcome function $g(s_1, \ldots, s_n): \times_{i=1}^n \mathcal{S}_i \mapsto \mathcal{Y}$. Together they define a mechanism $\Gamma= (\mathcal{S}_1, \ldots, \mathcal{S}_n, g(\cdot))$, which transforms users' strategies into a social choice function through the outcome function $g(\cdot)$. $(\Gamma, F, \lbrace u_i\rbrace_{i=1}^n)$ defines a Bayesian Game with payoffs $u_i(g(s_1, \ldots, s_n), \boldsymbol{\theta}_i)$ and strategies $\mathbf{s}_i:\boldsymbol{\Theta}_i\mapsto \mathcal{S}_i$.

The \textit{revelation principle} \cite{Osborne:1994aa} allows us to focus on direct mechanisms, i.e. $\mathcal{S}_i = \boldsymbol{\Theta}_i$ and $g(s_1, \ldots, s_N) \equiv g(\boldsymbol{\theta}) = f(\boldsymbol{\theta})$, which is the well-known fact that any equilibrium of any mechanism is identical to an equilibrium of a direct mechanism, provided truthful reporting. We focus on the dominant strategy equilibrium:

\begin{definition}[Dominant Strategy Equilibrium (DSE)]
A Dominant Strategy Equilibrium is given by
\begin{align}\label{eq:DSE_definition}
\hspace*{-0.2cm}\boldsymbol{\theta}_i = \arg\max_{\mathbf{z}_i\in\boldsymbol{\Theta}_i} \mathbb{E}_{\mathbf{z}_i}\left[u_i(f(\mathbf{z}_i, \mathbf{z}_{-i}), \boldsymbol{\theta}_i)\right]\quad\forall i\in\mathcal{I},~ \mathbf{z}\in\boldsymbol{\Theta}
\end{align}
That is, if the supremum of user $i$'s expected utility $u_i$ is achieved with truthful reporting $s_i^\ast(\boldsymbol{\theta}_i) = \boldsymbol{\theta}_i$, regardless of other users reports $\mathbf{z}_{-i}\in\boldsymbol{\Theta}_{-i}$, then the social choice function $f(\cdot)$ is \textit{dominant strategy incentive compatible}.
\end{definition}

\subsection{Timing, User Types, and Reward Calculation}
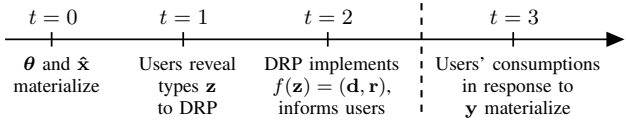
\begin{figure}[h]
\centering
\resizebox{\linewidth}{!}{

\begin{tikzpicture}
\draw[line width=0.25mm,->] (0,0) -- (9,0);
\foreach \x in {1.15, 4.4, 7.97, 12.55}{
   \draw [line width=0.25mm](\x/1.7,3pt) -- (\x/1.7,-3pt);
}

\draw [line width=0.3mm, dashed](6.05,15pt) -- (6.05,-30pt);

\draw (0.7,0) node[below=3pt] {\begin{footnotesize}
\begin{tabular}{c}
	$\boldsymbol{\theta}$ and $\mathbf{\hat{x}}$\\
    materialize
\end{tabular}
\end{footnotesize}} node[above=3pt] {{\small $t=0$}};

\draw (2.6,0) node[below=3pt] {\begin{footnotesize}
\begin{tabular}{c}
	Users reveal\\
    types $\mathbf{z}$\\
    to DRP
\end{tabular}
\end{footnotesize}} node[above=3pt] {{\small $t=1$}};

\draw (4.7,0) node[below=3pt] {\begin{footnotesize}
\begin{tabular}{c}
    DRP implements\\
    $f(\mathbf{z}) = (\mathbf{d}, \mathbf{r})$,\\
    informs users
\end{tabular}
\end{footnotesize}} node[above=3pt] {{\small $t=2$}};

\draw (7.4,0) node[below=3pt] {\begin{footnotesize}
\begin{tabular}{c}
    Users' consumptions\\
    in response to\\
    $\mathbf{y}$ materialize
\end{tabular}
\end{footnotesize}} node[above=3pt] {{\small $t=3$}};

\end{tikzpicture}
}
\caption{DR Mechanism Timeline}
\label{fig:timeline}
\end{figure}

The DR mechanism unfolds as follows:
\begin{itemize}
\item The users $i\in\mathcal{I}$ discover their types $\boldsymbol{\theta}_1, \ldots, \boldsymbol{\theta}_n$. The baselines $\hat{x}_1, \ldots, \hat{x}_n$ become common knowledge. 

\item The users reveal their types $\lbrace \mathbf{z}_i\rbrace_{i=1}^n$ to the DRP, where $\mathbf{z}_i$ not necessarily corresponds to the true type $\boldsymbol{\theta}_i$.
\item The DRP implements the collective choice $f(\mathbf{z}) = \mathbf{y} = (\mathbf{d}, \mathbf{r})$ through the mechanism $\Gamma$.
\item Users observe $f(\mathbf{z})$ and adjust their consumption according to \eqref{eq:semilog_demand_curve} and $d_i, r_i$.
\end{itemize}
For better visualization, Figure \ref{fig:timeline} depicts these steps.

An important observation is that, after the implementation of $f(\mathbf{z})$ at $t=2$, the DRP calculates its \textit{expected} profit $\mathbb{E}[\Pi]$ and the \textit{expected} payments disbursed to each user $i$. Due to the Myerson-Satterthwaite Theorem \cite{Myerson:1983aa}, we do not perform any ex-post analysis on the \textit{realized} consumptions $\mathbf{x}(\mathbf{r})$ at $t=3$.

To model the fact that users' base electricity consumption is often driven by habits rather than rational profit-maximization \cite{Marechal:2010aa}, we assume the user-specific intercept $\bar{x}_i$ to be drawn from an a-priori defined distribution $G$ with characteristic parameters $\boldsymbol{\xi}_i$ encoded in user $i$'s private type. $\boldsymbol{\xi}_i$ itself is distributed according to the joint distribution $F_{\boldsymbol{\xi}}$, and so $\bar{x}_i$ is a compound random variable. The slope, however, is assumed to be explicitly known for each user and drawn from distribution $F_{\alpha}$. Thus $\boldsymbol{\theta}_i = (\alpha_i\sim F_{\alpha}, \boldsymbol{\xi}_i\sim F_{\boldsymbol{\xi}})$, where $\bar{x}_i\sim G_{\boldsymbol{\xi}_i}\sim G_{\boldsymbol{\xi}_i\sim F_{\boldsymbol{\xi}}}$. All distributions have support on $\mathbb{R}_+$. We make the following assumption:


\begin{assumption}\label{as:independent_random_variables}
The types $(\alpha_i, \boldsymbol{\xi}_i)$ are drawn from independent, absolutely continuous distributions $F_{\alpha}$ and $F_{\boldsymbol{\xi}}$. Each component $k$ in $\boldsymbol{\xi}_i$ is independently drawn from the marginal distribution $F_{\boldsymbol{\xi}_k}$ s.t. $F_{\boldsymbol{\xi}} = F_{\boldsymbol{\xi}_1}\cdot\ldots\cdot F_{\boldsymbol{\xi}_m}$, where $m$ is the dimension of $\boldsymbol{\xi}_i$. $G$ is pairwise independent of $F_{\alpha}$ and $F_{\boldsymbol{\xi}}$.
\end{assumption}

User $i$'s expected utility $\mu_i$, given the realized types $\alpha_i$ and $\boldsymbol{\xi}_i$, allocation $d_i=1$, and reward $r_i$, is obtained by taking the expectation of \eqref{eq:utility_under_dr_mechanism} with respect to the random variable $\bar{x}_i\sim G_{\boldsymbol{\xi}_i}$:
\begin{align}\label{eq:user_expected_profit}
\mu_i(d_i=1, r_i) = \int_{\mathbb{R}_+}u_i(\alpha_i, r_i, x)~ dG_{\boldsymbol{\xi}_i}(x),
\end{align}
which is strictly monotonically increasing in reward $r_i$, cf. \eqref{eq:utility_under_dr_mechanism}. Letting $\mathcal{G}$ denote the CDF of $G$, \eqref{eq:user_expected_profit} for $r_i = 0$ becomes
\begin{align*}
\mu_i(d_i=1, r_i=0) = q_i\left[\hat{x}_i(1-\mathcal{G}(\hat{x}_i)) - \int_{\hat{x}_i}^\infty x ~dG_{\boldsymbol{\xi}_i}(x) \right]
\end{align*}
which is negative. Hence, there is a unique $\tilde{r}_i$ such that $\mu_i(d_i=1, \tilde{r}_i)=0$, i.e. the unique threshold reward level for which user $i$'s expected utility is zero. We approximate $\tilde{r}_i$ with Newton's method, exploiting the fact $\mu_i$ is monotonically increasing in $r_i$. Due to the same property, any reward $r_i \geq \tilde{r}_i$ fulfills the IR constraint as $\mu_i(d_i=0) = 0$ (Eq. \ref{eq:utility_under_dr_mechanism}).

\subsection{Mechanism for Demand Response}
We now present the Demand Response Mechanism:
\begin{enumerate}
\item Each user announces her private type $\mathbf{z}_i \in\boldsymbol{\Theta}_i$ to the DRP.  We will later show that this mechanism is incentive compatible, so that users report their types truthfully. In the following, we thus let $\mathbf{z}_i = \boldsymbol{\theta}_i$.
\item The DRP calculates the unique $\tilde{r}_i$ for each user based on the reports $\boldsymbol{\theta}_i$ with Newton's method on \eqref{eq:user_expected_profit}.
\item The DRP sorts $\lbrace \tilde{r}_i~|~i\in\mathcal{I}\rbrace$ in ascending order. Call this sorted set $\mathcal{R}$.
\item The DRP implements the social choice $\mathbf{y}$ as follows:
\begin{subequations}
\begin{align}
j_{\max} &= \min_j \left\{ j\in\mathbb{N}_+~\Big|~\sum_{i=1}^j \delta_i(\tilde{r}_j|\boldsymbol{\theta}_i) \geq M \right\}\label{eq:mechanism1_one}\\
j(i) &= \min_k \left\{ k\in\mathbb{N}_+ ~\Big|~ \sum_{s=1, s\neq i}^k \delta_s(\tilde{r}_k|\boldsymbol{\theta}_s)\geq M \right\}\nonumber\\
&\quad\quad\quad\quad\forall~ i\in\lbrace 1, \ldots, j_{\max}\rbrace=:\mathcal{T}\label{eq:mechanism1_two}\\
r_i &\leftarrow \tilde{r}_{j(i)} \geq \tilde{r}_i\quad\forall~ i\in\mathcal{T}\label{eq:mechanism1_three}
\end{align}
\end{subequations}
The allocation decision and the reward vector are
\begin{subequations}
\begin{align}
\mathbf{d} &= (1, \ldots, 1, \mathbf{0}_{n-j_{\max}}),\\
\mathbf{r} &= (\tilde{r}_{j(1)}, \ldots, \tilde{r}_{j(j_{\max})}, \mathbf{0}_{n-j_{\max}}).
\end{align}
\end{subequations}
\end{enumerate}
In the above mechanism, $\delta_i(\tilde{r}_j|\boldsymbol{\theta}_i)$ denotes the expected reduction of user $i$, given the reward level $\tilde{r}_j$ conditional on truthful reporting $\mathbf{z}_i = \boldsymbol{\theta}_i$, which is computed by taking the expectation on \eqref{eq:reduction_user} and \eqref{eq:semilog_demand_curve} with respect to $\boldsymbol{\xi}_i$.

The mechanism first determines the set of targeted users $\mathcal{T}$ by selecting the smallest index $j_{\max} \in\lbrace 1, \ldots, n \rbrace$, such that the sum of expected reductions of users $1$ through $j_{\max}$, if each user were given the reward $\tilde{r}_{j_{max}}$, exceeds the desired aggregate amount $M$ \eqref{eq:mechanism1_one}. Notice that since the set $\mathcal{R}$ is sorted in ascending order, $\tilde{r}_{j_{\max}}\geq\tilde{r}_i~\forall~i\leq j_{\max}$. Because $\mu_i(d_i=1, r_i)$ is strictly monotonically increasing in $r_i$, all targeted users will respond to incentive level $\tilde{r}_{j_{\max}}$.

Next, the reward for each user $i\in\mathcal{T}$ is determined by running the same exact mechanism \eqref{eq:mechanism1_one} on $\mathcal{I}\setminus i$, i.e. the set of all users excluding $i$ \eqref{eq:mechanism1_two}. Denote the user with the largest threshold reward $\tilde{r}_{j(i)}$ in this new set with $j(i)$. This reward level is then assigned to user $i$ \eqref{eq:mechanism1_three}.
 
In summary, the first $j_{\max}$ users \eqref{eq:mechanism1_one} with the smallest threshold rewards $\tilde{r}_i$ are offered user-specific unit-rewards (\eqref{eq:mechanism1_two}, \eqref{eq:mechanism1_three}). The remaining $n-j_{\max}$ users are not targeted.

Lastly, to ensure that the mechanism returns a valid index $j_{\max}$, we restrict $M$ to the range $\left[0,~ \sum_{i=2}^{n-1}\delta_i(\tilde{r}_{n-1}|\boldsymbol{\theta}_i)\right]$. If $M$ exceeds this range, there are not enough users to achieve expected aggregate reduction $M$ on the given $n$ users.

\begin{theorem}\label{thm:mech_1_incentive_compatible}
If $M\in \left[0,~ \sum_{i=2}^{n-1}\delta_i(\tilde{r}_{n-1}|\boldsymbol{\theta}_i)\right]$, the DR Mechanism terminates. The mechanism fulfills the IR constraint. Truthful reporting, i.e. $s_i^\ast(\boldsymbol{\theta}_i) = \boldsymbol{\theta}_i$, establishes a DSE.
\end{theorem}

Since truthful reporting establishes a DSE (Theorem \ref{thm:mech_1_incentive_compatible}), Mechanism I is also IC, due to the revelation principle \cite{Milgrom:2004aa}.

\begin{remark}
Due to the fact that $\lbrace (\alpha_i, \boldsymbol{\xi}_i)\rbrace_{i=1}^n$ are realizations of continuous random variables, no ties need to be broken in \eqref{eq:mechanism1_one}, \eqref{eq:mechanism1_two} and the sorting of the users into $\mathcal{R}$, because identical threshold rewards $\tilde{r}_i=\tilde{r}_j,~i,j\in\mathcal{I},~i\neq j$, only occur with probability zero.
\end{remark}
The presented mechanism runs in $\mathcal{O}(n\log n)$ time, as it takes $\mathcal{O}(n\log n)$ time to create the sorted list $\tilde{R}$ and $\log n$ time to determine the correct index $j_{\max}$ \eqref{eq:mechanism1_one} with a binary search on all possible values of $j=1, \ldots, n$. Once $j_{\max}$ has been found, we have to determine the reward level for each user by running the same mechanism again, which amounts to $\mathcal{O}(n\log n)$. This yields a runtime of $\mathcal{O}(n\log n)$.
\begin{remark}
This mechanism is motivated by the classic Vickrey-Clarke-Groves Mechanism \cite{Milgrom:2004aa}, as it allocates an ``items'' (in our case reward) to the ``highest'' bidders (in our case lowest threshold reward levels).
\end{remark}

\subsection{Numerical Example}
Table \ref{tab:Example_Part1} lists threshold rewards $\tilde{r}_i$ and reduction functions of $6$ hypothetical users in a synthetic user pool. The linearity of $\lbrace \delta_i\rbrace_{i=1}^6$ is assumed for ease of exposition. Let $M=4.3$.
\begin{table}[h]
\centering
\begin{tabular}{*7c}
\toprule
\multicolumn{7}{c}{Pool of Users} \\
\hline
User$\#$ & 1 & 2 & 3 & 4 & 5 & 6\\
\hline
$\tilde{r}_i$ & 0.5 & 1.0 & 1.5 & 1.8 & 2.0 & 2.1 \\
$\delta_i(r_i)$ & $1+r_1$ & $2 + \frac{r_2}{2}$ & $1+\frac{r_3}{3}$ & $2+\frac{r_4}{4}$ & $1+\frac{r_5}{2}$ & $1+\frac{r_6}{5}$ \\
\bottomrule
\end{tabular}
\vspace{0.1cm}
\caption{Example User Characteristics}
\label{tab:Example_Part1}
\end{table}

\eqref{eq:mechanism1_one} selects $j_{\max}=2$ such that $\delta_1(\tilde{r}_2) + \delta_2(\tilde{r}_2) = (1+1)+(2+\frac{1}{2}\cdot 1) = 4.5 \geq M$. Thus $\mathcal{T}=\lbrace 1, 2\rbrace$. \eqref{eq:mechanism1_two} then determines $j(1)$ and $j(2)$ by solving \eqref{eq:mechanism1_one} on $\mathcal{T}\setminus 1$ and $\mathcal{T}\setminus 2$, respectively:
\begin{itemize}
\item For $i=1$, $j(1)=4$ because $\delta_2(\tilde{r}_4) + \delta_3(\tilde{r}_4) + \delta_4(\tilde{r}_4) = (2+1.8/2)+(1+1.8/3)+(2+1.8/4) = 6.95 \geq M$. Indeed, $j(1)\neq 3$ because $\delta_2(\tilde{r}_3)+\delta_3(\tilde{r}_3) = (2+1.5/2)+(1+1.5/3)=4.25 < M$.
\item For $i=2$, $j(2)=4$ because $\delta_1(\tilde{r}_4) + \delta_3(\tilde{r}_4) + \delta_4(\tilde{r}_4) = (1+1.8)+(1+1.8/3)+(2+1.8/4) = 6.85 \geq M$. Indeed, $j(2)\neq 3$ because $\delta_1(\tilde{r}_3)+\delta_3(\tilde{r}_3) = (1+1.5)+(1+1.5/3)=4 < M$
\end{itemize}
User 1 and 2's rewards therefore are $\tilde{r}_4$, see \eqref{eq:mechanism1_three}.

\section{Effect of Baseline ``Gaming''}
\label{sec:baseline_gaming}

By expanding user $i$'s reduction of consumption \eqref{eq:reduction_user},
\begin{align}\label{eq:reduction_components}
\delta_i = (\hat{x}_i - \bar{x}_i) + \bar{x}_i(1-e^{-\alpha_i r_i}) =: \delta_i^{\text{BL}} + \delta_i^{r},
\end{align}
it becomes clear that the measured reduction $\delta_i$ of user $i$ is comprised of two components: $\delta_i^{\text{BL}}$, which captures the difference between the baseline $\hat{x}_i$ and the base consumption (i.e. the consumption with no reward), and the actual reduction $\delta_i^r$ due to the elasticity of user $i$ in response to the reward level $r_i$. $\delta_i^{\text{BL}}$ is a ``virtual reduction'', which, if positive (negative), represents the amount of falsely measured reduction (increase). From an economic perspective, $\delta_i^{\text{BL}} > 0$ results in falsely allocated credit from the utility to the DRP as well as from the DRP to users $i$. On the contrary, $\delta_i^{\text{BL}} < 0$ is synonymous with a misallocated monetary transfer from user $i$ to the utility as well as from the utility to the DRP proportional to the amount of $|\delta_i^{\text{BL}}|$. To diminish the effect of virtual reduction, the baseline estimates should become as precise as possible.
We make the following assumption:
\begin{assumption}\label{as:independent_random_variables_time}
The random variables $\alpha_i$ and $\boldsymbol{\xi}_i$ for different points in time are independent.
\end{assumption}
Assumption \ref{as:independent_random_variables_time} excludes the possibility of baseline manipulation \cite{Campaigne:2016aa}, which captures the fact that users can inflate or deflate their baseline, given the knowledge of future DR events, in order to increase their calculated reduction $\delta_i$ \eqref{eq:reduction_user}. For example, a user can increase her expected utility \eqref{eq:utility_under_dr_mechanism} for a DR event by consciously over-consuming prior to the DR event so as to increase the baseline $\hat{x}_i$, which results in a higher payment $r_i\cdot[\hat{x}_i - x_i]_+$, despite having a zero actual reduction $\delta_i^r$. However, as DR events are difficult to forecast, the mild assumption that users do not consciously manipulate their baseline justifies Assumption \ref{as:independent_random_variables_time}, that is, users consume independently of the past and the future.

As a result, averaging 10 recent observations for weekdays (or 4 for weekends and holidays), excluding hours of past DR events, results in an unbiased estimate of the mean consumption $x_i$, but with considerable variance around $x_i$. From a theoretical perspective, the baseline estimate approaches zero variance as the number of previous observations to estimate $\hat{x}_i$ goes to infinity, due to the Central Limit Theorem and Assumption \ref{as:independent_random_variables_time}. In the next Section, we simulate the effect of more precise baseline estimates on the quantity of virtual reductions $\delta_i^{\text{BL}}$.

As the analysis of the economic implications of this virtual baseline reduction component is outside the scope of the paper, the reader is referred to \cite{Borenstein:2002aa}, which explicitly characterizes the magnitude of marginal competitive rents in California's wholesale electricity market, and \cite{Zhou:2016aa, Zhou:2016ab}, where the authors suggest alternative baselining methodologies based on Machine Learning, which weaken the effect of such virtual reductions.

%
%


\section{Simulations}
\label{sec:Simulations}
In this section, we simulate the presented mechanism and the effect of virtual reductions stemming from imperfect baseline predictions. We utilize hourly smart meter data from 1,000 residential customers serviced by the three largest utilities in California (Pacific Gas \& Electric, San Diego Gas \& Energy, and Southern California \& Edison).

\subsection{Approximation of Base Consumption}
\begin{figure}[h]
\centering
\includegraphics[scale=0.35]{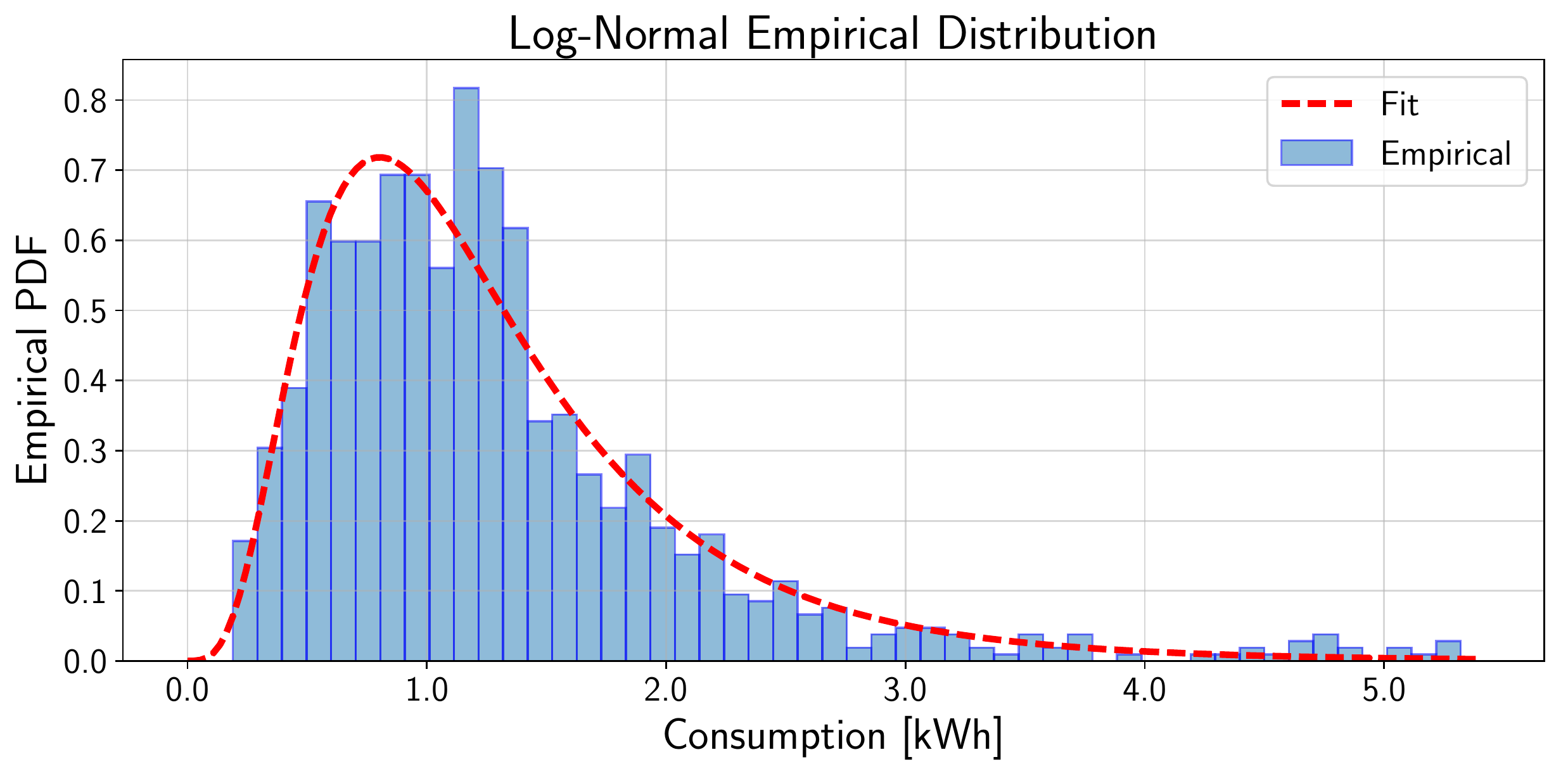}
\caption{Lognormal Consumption Distribution Fit for Selected User, 5-6 pm}
\label{fig:lognormal_consumption_distribution}
\end{figure}

Figure \ref{fig:lognormal_consumption_distribution} shows the distribution of the hourly base consumptions between 5-6 pm in the absence of DR events of a selected user. The restriction to 5-6 pm is arbitrarily chosen. For a more thorough analysis, we would have to analyze all 24 hours of the day separately.

It is found that the base consumption $\bar{x}_i$ can be approximated with a log-normal distribution, whose density
\begin{align}\label{eq:lognormal_distribution}
\mathcal{N}(\log x) = \frac{1}{\sigma \sqrt{2\pi}}\exp\left( -\frac{(\log (x-\ell)-\mu)^2}{2\sigma^2} \right)
\end{align}
is fully parameterized by the shape $\sigma > 0$, scale $s = e^\mu > 0$, and location parameter $\ell$. As \eqref{eq:lognormal_distribution} has support on $(\ell, \infty)$, the location $\ell$ denotes the lower bound on the support of the base consumption distribution. 

\begin{figure*}[hbtp]
{\includegraphics[width=1.0\textwidth]{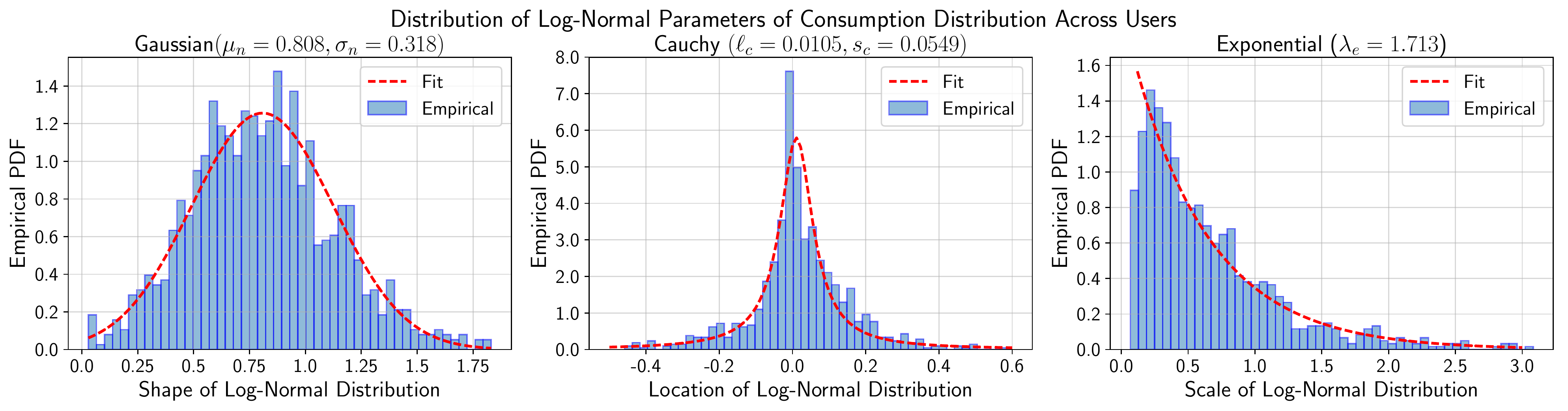}} \hfill \vspace*{-0.2cm}
\caption{Compound Statistics for Lognormal Consumption Distribution. Left: Shape, Middle: Location, Right: Scale}
\label{fig:compound_statistics}
\vspace*{-0.2cm}
\end{figure*}

Fitting a log-normal distribution to the hourly consumptions between 5-6 pm across all users yields a distribution of the compound statistics $\boldsymbol{\xi}_i=(\sigma,~s,~\ell)$, given below: 
\begin{align*}
\bar{x}_i &\sim \mathrm{Lognormal}(\sigma, s, \ell) \quad\quad &\sigma &\sim\mathcal{N}(\mu_{n}, \sigma_{n}) \\
s &\sim\mathrm{Cauchy}(\ell_c, s_c) \quad\quad
&\ell &\sim\mathrm{Exponential}(\lambda_e)
\end{align*}
That is, the shape parameter $\sigma$ is best approximated with a Gaussian distribution $\mathcal{N}(\mu_{n}, \sigma_{n})$, the location $\ell$ by a Cauchy distribution parameterized by location $\ell_c$ and scale parameter $s_c$, and the scale parameter $s$ by an exponential distribution with parameter $\lambda_e$. Figure \ref{fig:compound_statistics} shows the distribution of these compound statistics across all 1,000 users.

\subsection{Performance of DR Mechanism}
We compare the DR Mechanism \eqref{eq:mechanism1_one}-\eqref{eq:mechanism1_three} to the hypothetical case of an omniscient DRP, which knows $\lbrace (\alpha_i, \boldsymbol{\xi}_i)\rbrace_{i=1}^n$. Despite this being an unrealistic scenario, it provides a near-optimal approximation of the minimum payment disbursed to the users necessary to elicit a target reduction of $M$. Given the sorted list $\mathcal{R}$ of user-specific threshold rewards, the omniscient DRP implements the social choice $\mathbf{y}^o = (\mathbf{d}^o, \mathbf{r}^o)$ as follows:

\begin{subequations}
\begin{align}
j^o &= \min_{j}\left\{ j\in\mathbb{N}_+\Big|\sum_{i=1}^j \delta_i(\tilde{r}_i)\geq M \right\}\label{eq:mechanism_omn_one}\\
\mathcal{T}^o &= \lbrace 1, \ldots, j^o\rbrace\label{eq:mechanism_omn_two}\\
r_i^o &= \tilde{r}_i\quad\forall~ i\in\mathcal{T}^o \label{eq:mechanism_omn_three}\\
\mathbf{d}^o &= (1, \ldots, 1, \mathbf{0}_{n-j^o})\label{eq:mechanism_omn_four}
\end{align}
\end{subequations}
That is, the DRP determines the smallest index $j^o$ to obtain the desired expected aggregate reduction $M$ \eqref{eq:mechanism_omn_one} where each user $\lbrace 1, \ldots, j^o\rbrace$ is given their individual threshold reward $\tilde{r}_i$ \eqref{eq:mechanism_omn_three}. These are the targeted users \eqref{eq:mechanism_omn_two}, \eqref{eq:mechanism_omn_four}.

Due to $\lbrace (\alpha_i, \boldsymbol{\xi}_i)\rbrace_{i=1}^n$ being publicly known, users are unable to extract information rent from the DRP, which is the payment to the users to elicit their private information \cite{Laffont:2002aa}. Hence, the DRP can offer targeted users their threshold reward $\tilde{r}_i$, which keeps users at an expected utility \eqref{eq:user_expected_profit} of zero. To guarantee user participation, the DRP has to offer the reward level $\tilde{r}_i + \varepsilon$ to each user $i\in\mathcal{T}^o$, where $\varepsilon$ is an arbitrarily small positive number.

\begin{figure}[hbtp]
\centering
\includegraphics[scale=0.285]{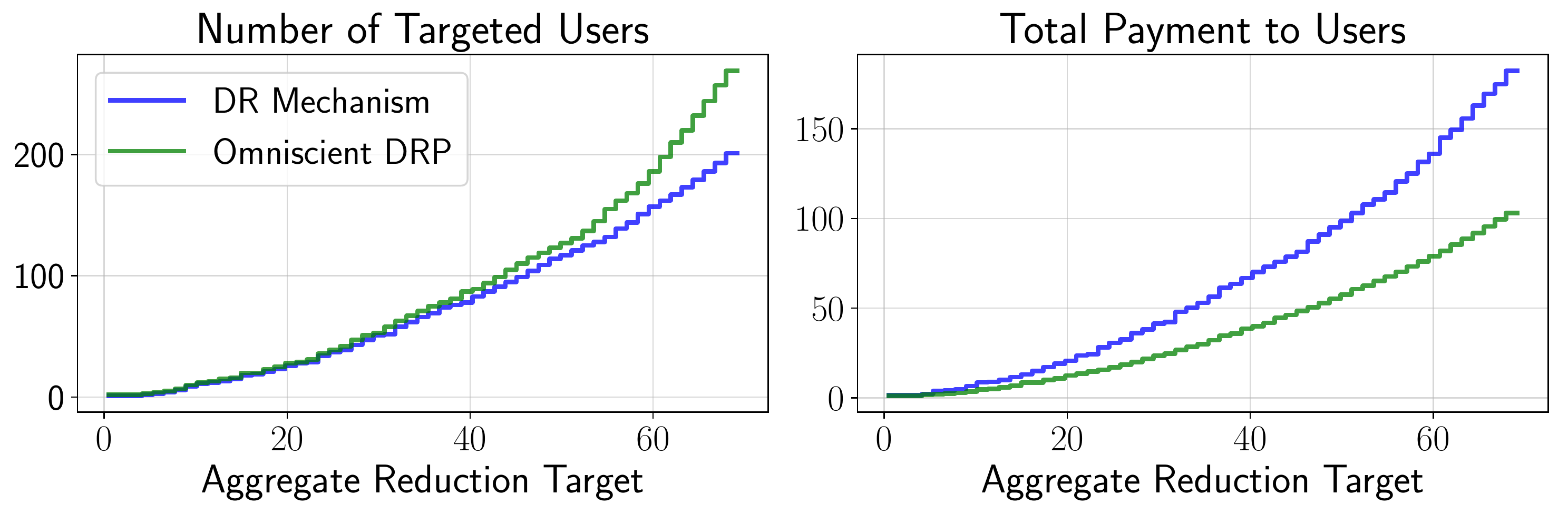}
\vspace*{0.1cm}
\caption{Number of Targeted Users and Total Payment to Users for DR Mechanism (blue) vs. Omniscient Allocation (green), $n=500,~q=5.0,~\alpha_i\sim\text{unif}[0.05, 0.06].$}
\label{fig:Comparison_Mechanisms}
\end{figure}

Figure \ref{fig:Comparison_Mechanisms} compares the DR Mechanism \eqref{eq:mechanism1_one}-\eqref{eq:mechanism1_three} to the omniscient allocation with respect to the number of targeted users (left) and the total amount of rewards disbursed (right) on $n=500$ users whose parameters $\boldsymbol{\xi}_i = (\sigma_i, s_i, \ell_i)$ are sampled from the fitted distributions in Figure \ref{fig:compound_statistics}. As expected, the omniscient allocation is more economical at eliciting a particular aggregate reduction target $M$ due to the lack of private user information, namely about $45\%$ better than the DR mechanism. However, it needs to target more customers as each customer in the omniscient case receives a smaller reward level than in the DR mechanism.

%

\subsection{Virtual Reductions}
\begin{figure}[hbtp]
\centering
\includegraphics[scale=0.288]{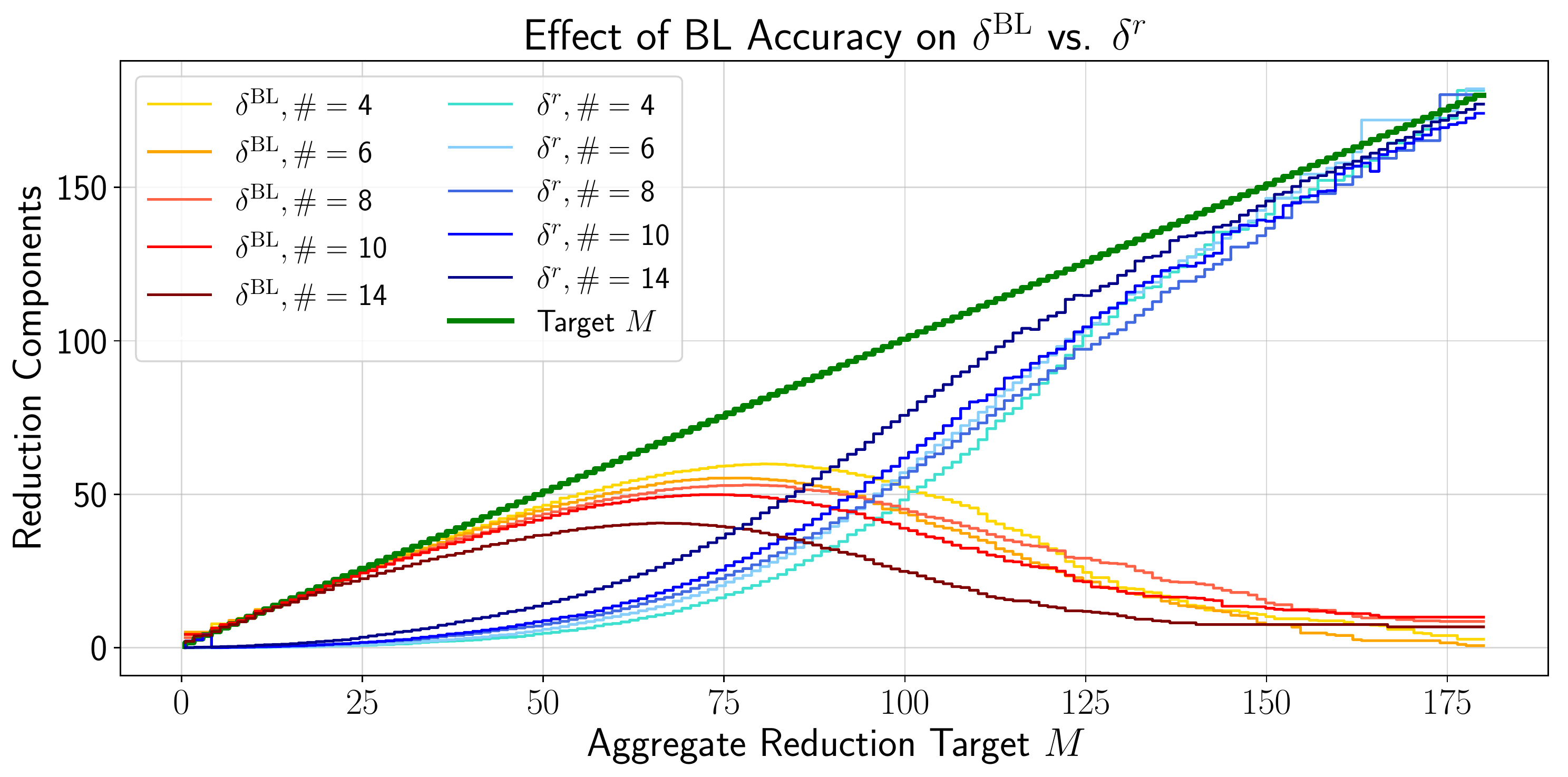}
\caption{Composition of Target Aggregate Reduction $M$ for varying Baselines. Red: $\sum_{i\in\mathcal{T}}\delta_i^{\text{BL}}$. Blue: $\sum_{i\in\mathcal{T}}\delta_i^r$. Parameters: $n=500, q=5.0, \alpha_i\sim\text{unif}~[0.05, 0.06]$}
\label{fig:BL_vs_actual_reduction}
\vspace*{-0.2cm}
\end{figure}

Figure \ref{fig:BL_vs_actual_reduction} shows the total reduction $\sum_{i\in\mathcal{T}}\delta_i$ of all targeted users and its components $\sum_{i\in\mathcal{T}}\delta_i^{\text{BL}}$ and $\sum_{i\in\mathcal{T}}\delta_i^{r}$ as a function of $M$ for $n=500$ users, $q=5$, and elasticities $\lbrace\alpha_i\rbrace_{i=1}^n$ drawn from a uniform distribution with support $[0.05, 0.06]$. The baseline computed with a particular number $x$ of previous days taken into consideration is calculated as the mean of $x$ randomly drawn samples from the empirical consumption distribution \eqref{eq:lognormal_distribution}. 

As can be seen from Figure \ref{fig:BL_vs_actual_reduction}, almost the entire reduction is attributed to the baseline component $\sum_{i\in\mathcal{T}}\delta_i^{\text{BL}}$ for small $M$. With larger values of $M$, the contribution of $\sum_{i\in\mathcal{T}}\delta_i^{\text{BL}}$ decreases marginally and finally starts decreasing. This can be explained by the fact that sorting users in $\mathcal{R}$ tends to put users with the highest $\delta_i^{\text{BL}}$ towards the start of the array, while those with the lowest (and negative) $\delta_i^{\text{BL}}$ bunch up at the end of $\mathcal{R}$. Consequently, as more users are assigned to $\mathcal{T}$, the sum of baseline reductions decreases. The actual reduction $\sum_{i\in\mathcal{T}}\delta_i^r$ increases exponentially with the number of users targeted, because as more users are assigned to $\mathcal{T}$, the per-unit reward levels also increase, which results in a superlinear growth of $\sum_{i\in\mathcal{T}}\delta_i^r$.

For increasing numbers of baseline averaging components, that is, the number of previous days to calculate the baseline, the variance of the baseline estimate $\bar{x}_i-\hat{x}_i$ decreases, and so the virtual reductions decrease. For the limiting case of a perfect baseline, the virtual reductions are zero.

\begin{figure}[hbtp]
\centering
\includegraphics[scale=0.288]{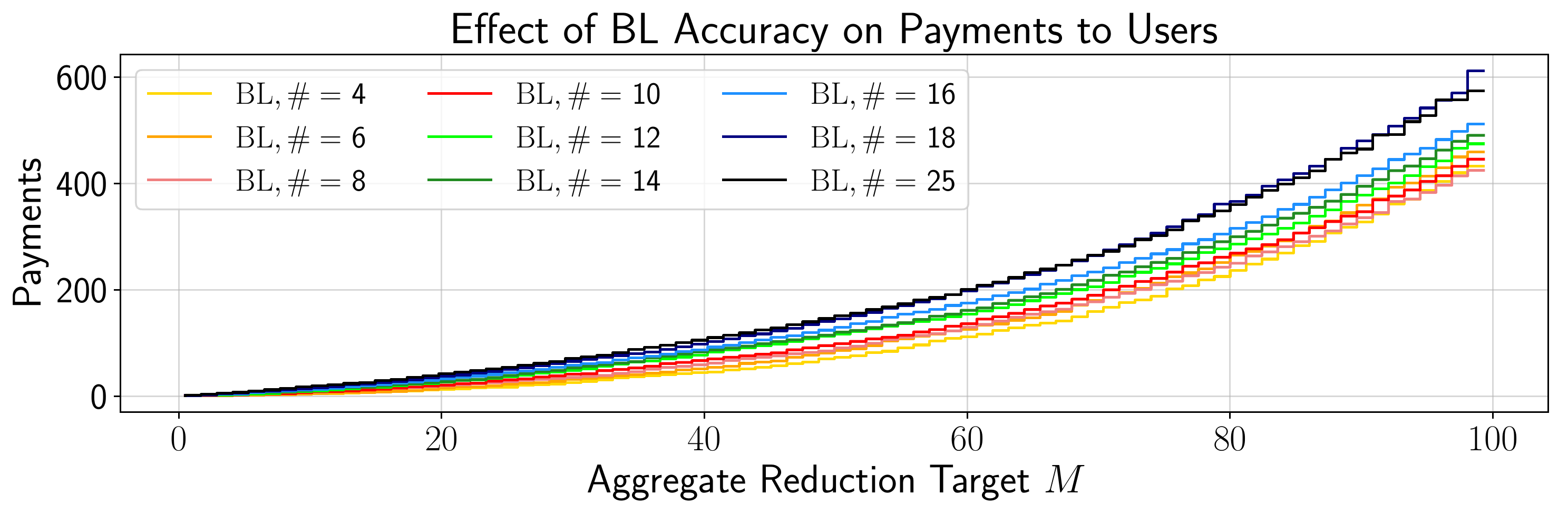}
\caption{Payments to Users to Elicit $M$ for varying Baseline Accuracies.}
\label{fig:BL_vs_total_payment}
\vspace*{-0.2cm}
\end{figure}

Finally, Figure \ref{fig:BL_vs_total_payment} depicts the total amount of payments the DRP has to make to the users for varying baseline accuracies in the range $M\in[0, 100]$, where virtual payments have the largest effect (see Figure \ref{fig:BL_vs_actual_reduction}). For more inaccurate baselines (fewer number of averaging days), the DRP has to pay users less as it can exploit the virtual reduction component.

%
%


\section{Conclusion}
\label{sec:Conclusion}
We modeled Residential Demand Response with a Mechanism Design Framework where a Demand Response Provider asks a subset of its customers under contract to reduce electricity consumption temporarily in exchange for a monetary reward. Each user's consumption in response to a per-unit reduction incentive is modeled as a logarithmic demand curve where the intercept and the slope are private information of users. While each user has a fixed slope, the user-specific intercept, which corresponds to the consumption given no incentive, is modeled as a realization of a compound random variable, capturing the fact that users often do not consume electricity in a profit-maximizing fashion, but rather are following habits, and hence have no explicit utility function. To make an informed choice about the magnitude of reductions in response to incentives to achieve an a-priori defined aggregate reduction target $M$, the Demand Response Provider asks for residential customers' bids to elicit their private information. Reductions are measured against a counterfactual estimate of the consumption in the hypothetical case of no DR event, which in this paper is the ``10-in-10''-baseline employed by the California Independent System Operator. Since this baseline is plagued by high variance, the Demand Response Provider can exploit ``virtual reductions'' emanating from high baseline estimates, which are false-positive reductions despite the users not having reduced, but whose role diminishes as the baseline becomes more precise. The Demand Response mechanism is validated on hourly smart meter data of residential customers in California.

Our analysis is an initial step towards quantifying economic implications of Demand Response on a residential level. While we approximated users' base demand (i.e. in the absence of incentives) reasonably well with existing smart meter data, the price elasticity of users in response to incentives is unknown, a fact that is complicated by the fundamental problem of causal inference. Thus, to further validate our analysis on real data, credible parameters for users' slope of the demand curve would be necessary. 

Lastly, extending the single period analysis employed in this paper towards a dynamical problem, which allows for baseline manipulation of users, is a logical next step. Comparing the ``10-in-10''-baseline to improved baseline estimates obtained with Machine Learning techniques, which exploit serial correlation of consumption time series, would shed further light on the economics of Residential Demand Response.


%
%


\section*{Appendix}
\label{sec:Appendix}

\subsection*{Proof of Theorem \ref{thm:mech_1_incentive_compatible}}
\subsubsection*{Individual Rationality}
Notice first that each user $i\in\mathcal{T}$ is given the reward $\tilde{r}_{j(i)}$, where $j(i)\geq j_{\text{max}} \geq i.$ The first inequality is a consequence of \eqref{eq:mechanism1_two}, which for each $i\in\mathcal{T}$ re-runs \eqref{eq:mechanism1_one} on the subset of users $\mathcal{I}\setminus i$. Thus, to achieve the aggregate reduction $M$ on users $\mathcal{I}\setminus i$, where each user would be given the highest threshold reward of the targeted group, requires more users to be targeted than running the same mechanism on $\mathcal{I}$. Hence $j(i)\geq j_{\text{max}}$. The second inequality is due to the fact that $\tilde{R}$ is sorted in ascending order, which also implies
\begin{align*}
\mathbb{E}[u_i(\tilde{r}_{j(i)}|\boldsymbol{\theta}_i)] \geq \mathbb{E}[u_i(\tilde{r}_i|\boldsymbol{\theta}_i)] = 0.
\end{align*}
due to the monotonically increasing property of the expected utility in the reward.
Thus, participation in the mechanism and being assigned to $\mathcal{T}$ results in a non-negative expected utility, compared to a zero utility for non-participation. On the other hand, users $i\not\in\mathcal{T}$ receive a zero payment and so the expected utility is zero.

\subsubsection*{Incentive Compatibility}
To show that the DR mechanism is incentive compatible, first note that the reward level $r(i)$ for each $i\in\mathcal{T}$ is determined \textit{independently} of user $i$'s bid $\mathbf{z}_i$. For each $i\not\in\mathcal{T}$, user $i$ is not given a reward. To show IC, we must therefore iterate through the following two cases:
\begin{enumerate}
\item $i\in\mathcal{T}$ for $\mathbf{z}_i = \boldsymbol{\theta}_i$, i.e. user $i$ is assigned treatment with truthful reporting. This implies user $i$ is given reward $\tilde{r}_{j(i)}$, which results in a positive expected utility. Now suppose user $i$ had reported $\mathbf{z}_i \neq \boldsymbol{\theta}_i$. Then either the user is still assigned treatment, in which case her reward remains the same, or the user is not assigned treatment, in which case her reward reduces to zero. Thus, misreporting could lead to a zero utility when the user could have had a positive expected utility.
\item $i\not\in\mathcal{T}$ for $\mathbf{z}_i = \boldsymbol{\theta}_i$, i.e. user $i$ is outside the treatment group with truthful reporting. If user $i$ had reported $\mathbf{z}_i \neq \boldsymbol{\theta}_i$, then either the user is still outside the treatment group, which results in a zero utility, or the user is now in the treatment group. In the latter case, note that user $i$ is assigned reward $\tilde{r}_{j_{\text{max}}}$, as $j_{\text{max}}$ is exactly the solution to \eqref{eq:mechanism1_two}.
Finally, because $\tilde{r}_{j_{\text{max}}} < \tilde{r}_i$ (due to $i\not\in\mathcal{T}$), the expected utility turns negative. Thus, misreporting does not improve the expected utility, but could lead to a negative expected utility when the user could have had a zero utility.
\end{enumerate}
Combining the two cases above, it follows that misreporting the true type either yields a utility that is identical to or less than the utility in case of truthful reporting. Therefore the maximum expected utility is attained with truthful reporting, $\mathbf{z}_i = \boldsymbol{\theta}_i$, and so the DR mechanism is incentive compatible.

Lastly, to show that the mechanism terminates if $0 \leq M \leq \sum_{i=2}^{n-1}\delta_i(\tilde{r}_{n-1}|\boldsymbol{\theta}_i)$, simply notice that $j(i) \leq n~\forall~i\in\mathcal{T}$ \eqref{eq:mechanism1_two} because $\delta_1(\tilde{r}_k|\boldsymbol{\theta}_1) \geq \delta_i(\tilde{r}_k|\boldsymbol{\theta}_i),~i\leq k\leq j(i)$ due to the monotonically increasing property of \eqref{eq:reduction_user} and \eqref{eq:user_expected_profit}. Hence running the mechanism on $\mathcal{T}\setminus i$ always satisfies $M$.

\bibliographystyle{IEEEtran}
\bibliography{bibliography}


\end{document}